\documentclass[11pt,dvips]{article}
\usepackage{epsfig,times} 
\usepackage{picinpar}
\makeatletter
\renewcommand{\section}{\@startsection{section}{1}{0in}
	{0.4\baselineskip}{0.1\baselineskip}{\Large\bf}}
\renewcommand{\subsection}{\@startsection{subsection}{2}{0in}
	{0.25\baselineskip}{-\baselineskip}{\large\bf}}
\renewcommand{\subsubsection}{\@startsection{subsubsection}{3}{0in}
	{0.1\baselineskip}{-\baselineskip}{\normalsize\bf}}
\makeatother
\begin{document}

%
\thispagestyle{myheadings}
%
\markright{OG 4.4.06}
\begin{center}
{\LARGE \bf Use of Instrumented Water Tanks for the Improvement
            of Air Shower Detector Sensitivity}
\end{center}

\begin{center}
{\bf Anthony L. Shoup$^{1}$, for the Milagro Collaboration}\\
{\it $^{1}$Department of Physics and Astronomy, University of
California, Irvine, CA, 92697}\\
\end{center}

\begin{center}
{\large \bf Abstract\\}
\end{center}
\vspace{-0.5ex}
   Previous works have shown that water Cherenkov detectors have
superior sensitivity to those of scintillation counters as applied to
detecting extensive air showers (EAS).  This is in large part due to
their much higher sensitivity to EAS photons which are more than five
times more numerous than EAS electrons.  Large area water Cherenkov
detectors can be constructed relatively cheaply and operated reliably.
A sparse detector array has been designed which uses these types of
detectors to substantially increase the area over which the Milagro
Gamma Ray Observatory collects EAS information. Improvements to the
Milagro detector's performance characteristics and sensitivity derived
from this array and preliminary results from a prototype array currently
installed near the Milagro detector will be presented.
%

\vspace{1ex}

%
%
\section{Introduction}
\label{intro_sec}

The field of Very High Energy (VHE) gamma-ray astronomy has exploded in recent years,
mainly pushed by the development of more sensitive telescopes.  The emphasis has
been to lower energy thresholds, improve angular and energy resolutions and
most importantly hadronic cosmic ray background rejection.

Considerable efforts have also been made to develop telescopes which detect VHE
extensive air showers (EAS) which have secondaries that survive to ground level, such
as Milagro and the Tibet Array. If reasonable sensitivity at VHE energies can be
achieved with these detectors, they will offer powerful capabilities, such as full
overhead sky coverage both day and night regardless of weather and skylight
conditions.  This would allow much higher temporal coverage of sources that are
already known to be highly variable, such as Active Galactic Nuclei.

The Milagro detector is progressing toward reaching the necessary VHE sensitivity.
It is a large (60m x 80m x 8m) water pond instrumented with 723 8''
photomultiplier tubes (pmts) in two layers. These pmts detect the Cherenkov
light produced by EAS secondaries passing through the optically clear water.
Its high altitude (2650m) and sensitivity to both photonic and leptonic EAS
components give it an energy threshold such that for zenith traversing sources the
peak primary energy will be ~ 1 TeV. After calibrations it
will have good angular resolution and hadronic cosmic ray rejection (see McCullough 1999
for more details).

To improve the sensitivity of the current Milagro detector, 172
instrumented, large area ($ 5 m^2 $), water Cherenkov detectors (tanks) will be
deployed around the pond to effectively extend its active area.  As discussed
below, this will improve both the energy and angular resolution of Milagro and
increase its hadronic cosmic ray rejection, thus improving its overall VHE
sensitivity. It can also be used to increase Milagro's efficiency for detecting
EAS below 1.0 TeV which have core positions significantly away from the Milagro pond.

\section{Water Tank Detector \& Array}
\label{detector_sec}

The criteria for selecting a detector design that will improve the performance of EAS
experiments are: low cost and low maintence (a large ground area needs to be covered,
typically at a remote high altitude site), high sensitivity to EAS secondary particles,
and good timing and particle density resolution.  Previous works (Yodh 1996) showed
that water Cherenkov detectors have superior sensitivity to those of scintillation
counters for detecting EAS secondaries.  Thus the tank design proposed here satisfies
these design criteria, although the particle density resolution is somewhat poor.
On average the pmt signal is about 100 photoelectrons for a through-going vertical
muon.

Figure 1 displays a crossectional view of a tank showing the position of the top-mounted,
downward-looking 8'' pmt and the Tyvek-lined bottom, sides, and floating top.  This
position of the pmt gives a fairly uniform response across the full tank, although it
does degrade the timing resolution somewhat compared to a bottom mounted, upward
looking position.  Due to its active material, water, the
tank is sensitive to both the photonic and leptonic components of EAS as opposed to
plastic scintillator based detectors which are mainly sensitive to the leptonic
component.  The Tyvek lining provides a diffusivly reflective inner surface
with $ > 90\% $ reflectivity at the important wavelengths determined from convoluting
the Cherenkov photon spectrum and pmt quantum efficiency (wavelengths around 350 nm).

\begin{figwindow}[1,r,%
{\hbox{\centerline{\epsfig{file=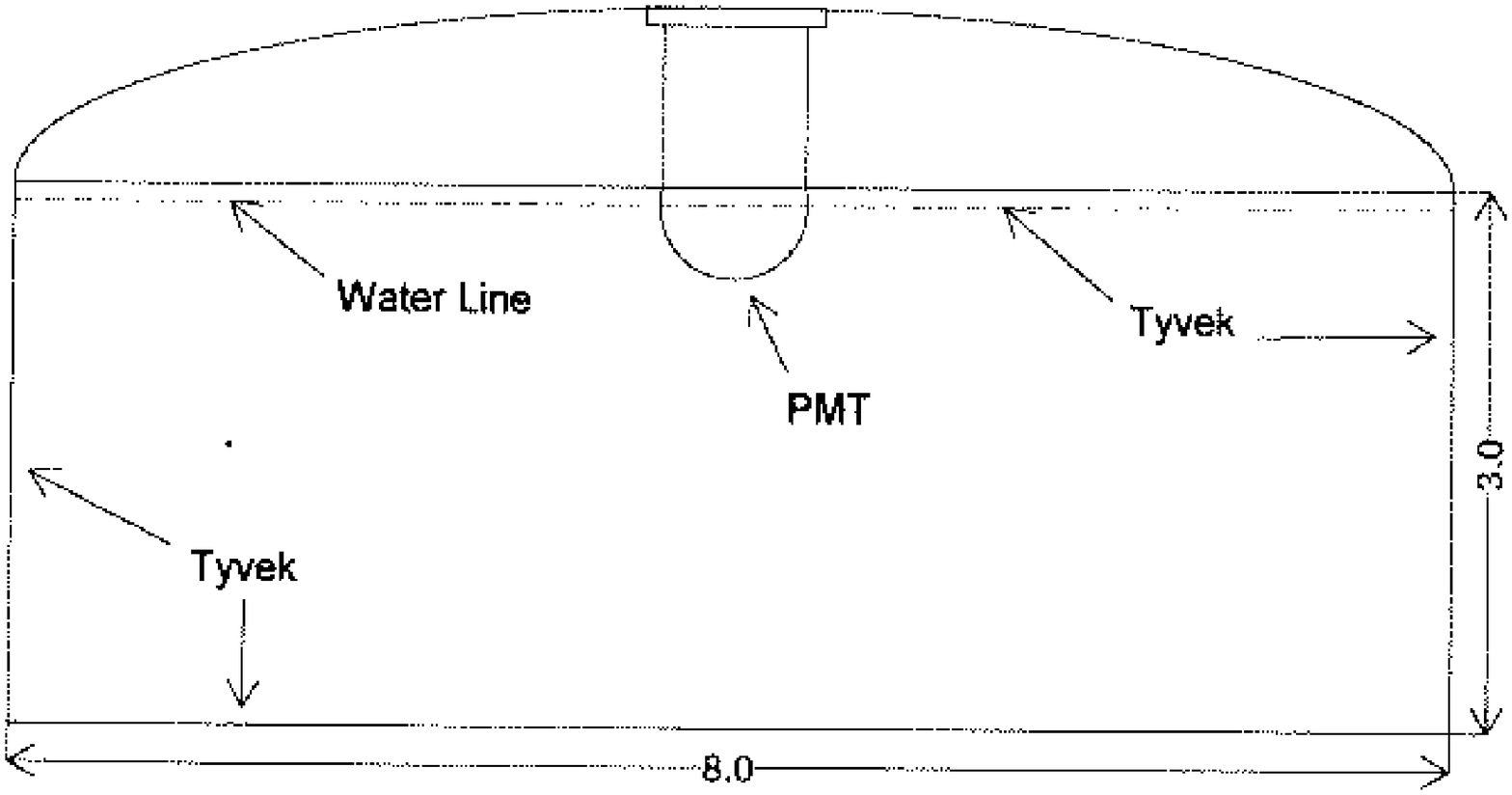,width=2.8in}}}},%
{Schematic of example water Cherenkov tank.  Key features are top mounted, downward
looking pmt and Tyvek lined inner surfaces (units are feet).}]
\end{figwindow}

The Milagro inspired tank array has 172 tanks placed on a square grid with a
spacing of 15 m, giving a full array area of 200m x 200m centered on the Milagro
pond.

Monte Carlo generated data was used to determine the performance characteristics of
these tanks, and the improvement of the sensitivity of the Milagro detector
generated by using these tanks.  Corsika was used for generating simulated EAS and
the Geant package was used to simulate the tank and Milagro detector responses (see
(Westerhoff 1998) for more details).

\section{Monte Carlo Estimates of Milagro Performance Improvements}

The information acquired with the tanks discussed above can be used in two separate
ways.  First, it may improve the angular and energy reconstruction resolutions of
EAS which trigger the Milagro pond detector by making additional independent
shower front timing measurements and by improving the EAS core position resolution for
EAS whose cores done not strike the pond.  A simple multiplicity trigger condition of
50 pond pmts being hit by an EAS was used as a pond-trigger in simulations.
Second, the information can be used to increase the effective area of the Milagro
detector by using it in a combined pond-tank trigger.

\subsection{Improvements in Pond Triggered Events}
\label{improve_pond_sec}

From simulation, on average about 24 tank pmts are hit per event where a hit is the
detection of 1 or more photoelectrons. The occupancy (fraction of the time a given pmt or
tank is hit) for pond pmts is about 30\% and for tanks is about 10\%.  The tanks
have fewer low pulse height hits than the pond pmts (below 30 pes) but about the
same number of large pulseheight hits (above 30 pes).

As seen in Figure 2, our simulations predict that using the tank array in
reconstructing EAS core positions can improve the position resolution
tremendously for EAS whose core positions are off
the pond. This improvement is crucial for EAS energy determination and is also
important in EAS angle determination because the pmt hit times must be corrected
for EAS shower front curvature about the core position. Current ongoing studies
of Monte Carlo generated EAS show that a good core position resolution should
improve the hadronic cosmic ray rejection capabilities of Milagro as well.

\begin{figwindow}[1,r,%
{\hbox{\centerline{\epsfig{file=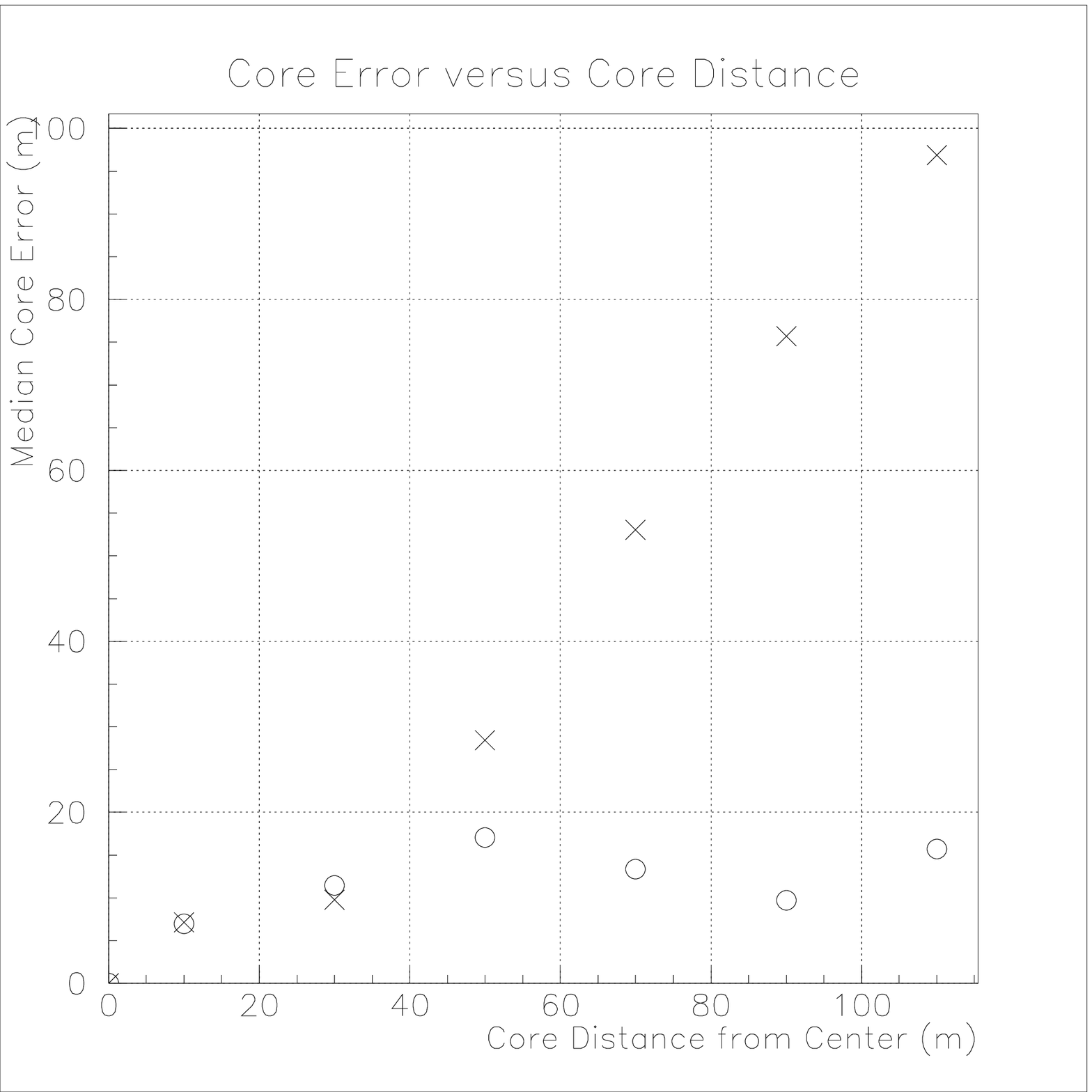,width=2.8in}}}},%
{Plot of median core position error versus core distance from center of Milagro pond
using tanks (circles) and not using tanks (crosses).}]
\end{figwindow}

Improvements in the angular reconstruction resolution of EAS is displayed in Figure
3.  The improvement is maximal for low multiplicity (number of pmts in Milagro
which detect light) events which are typically low primary energy EAS.  It is also
maximal for EAS whose cores land far from the pond.

\begin{figwindow}[1,r,%
{\hbox{\centerline{\epsfig{file=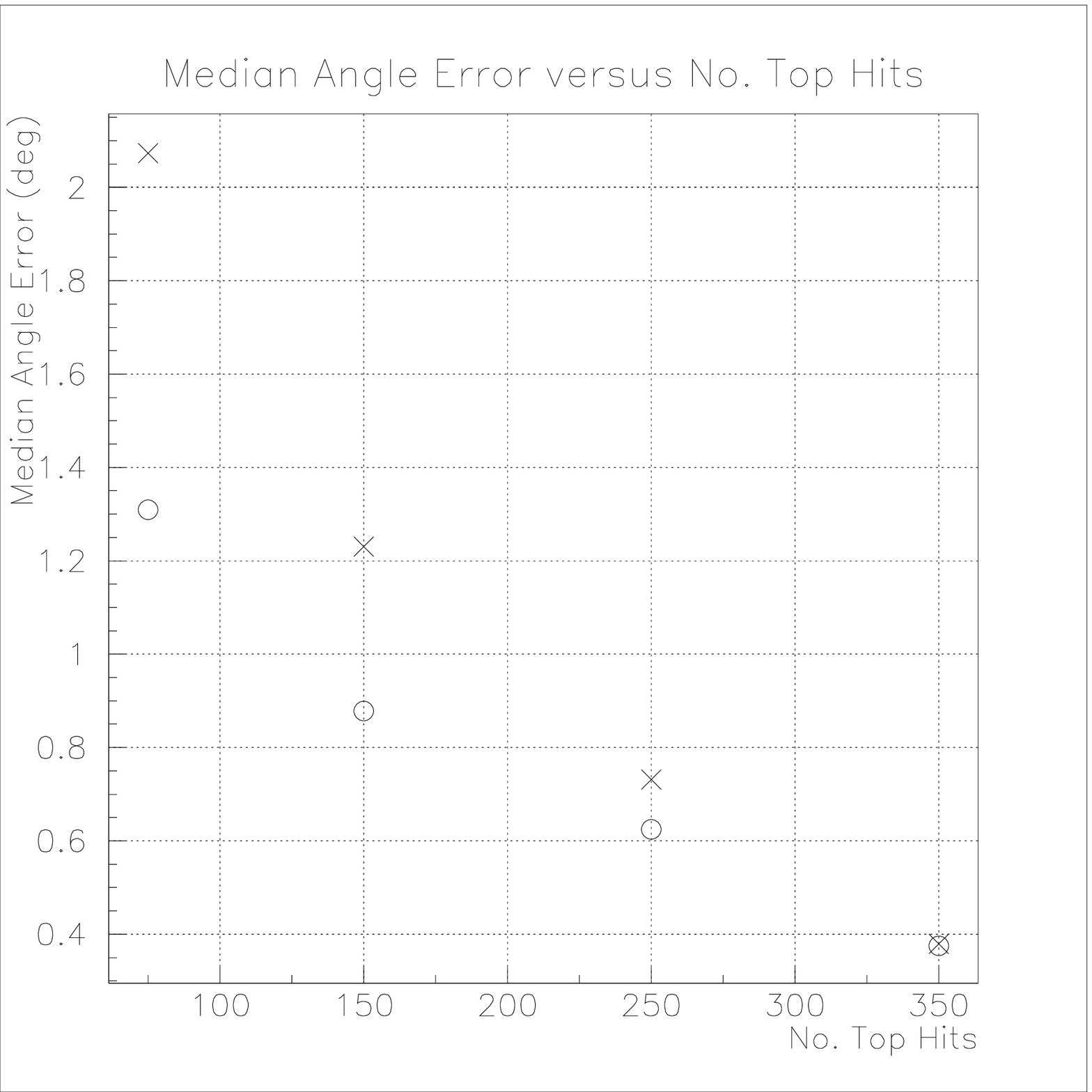,width=2.8in}
\epsfig{file=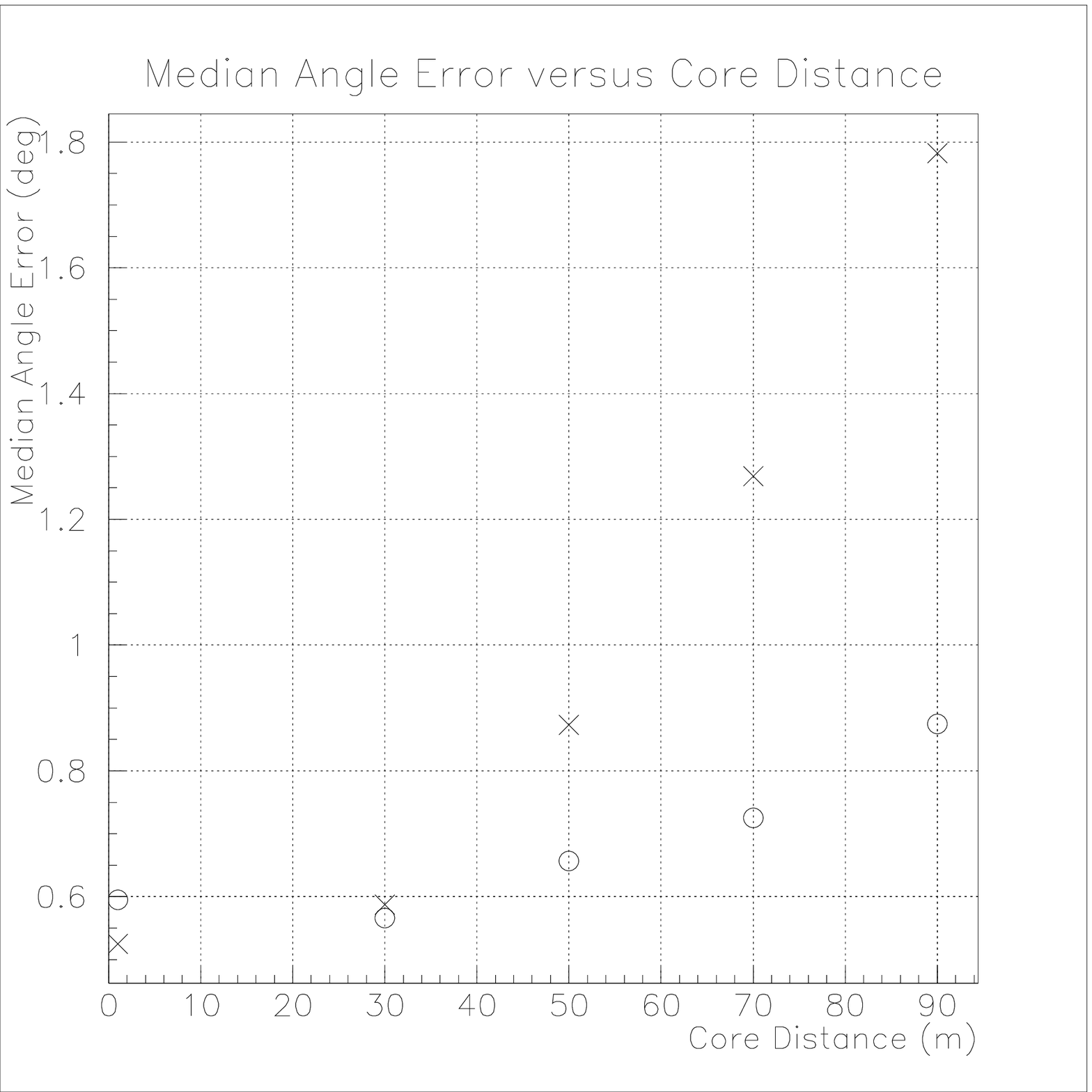,width=2.8in}}}},%
{Plots of median angle error versus number of pmt hits in top layer of Milagro pond
and versus EAS core distance from center of Milagro pond.  Circles are values when
tanks are used and crosses are value when tanks are not used.}]
\end{figwindow}

\subsection{Improvements in Trigger Sensitivity}
\label{improve_triger_sec}

Including tank acquired information within the Milagro trigger condition can increase
the efficiency for seeing low energy events.  This is clearly seen in Figure 4 which
displays a plot of the effective area of the Milagro detector for three types of triggers.
The pond-only trigger is a requirement that at least 50 pmts be hit by the EAS.  The
tank+pond trigger is that either the pond trigger be satisfied or that at least 5
tanks be hit by the EAS.  The tank-only trigger is that at least 5 tanks be hit and that
less than 50 pond pmts be hit by the EAS.  The tank-only trigger is included to
explicitly show the contribution of the tanks to the effective area.

\begin{figwindow}[1,r,%
{\hbox{\centerline{\epsfig{file=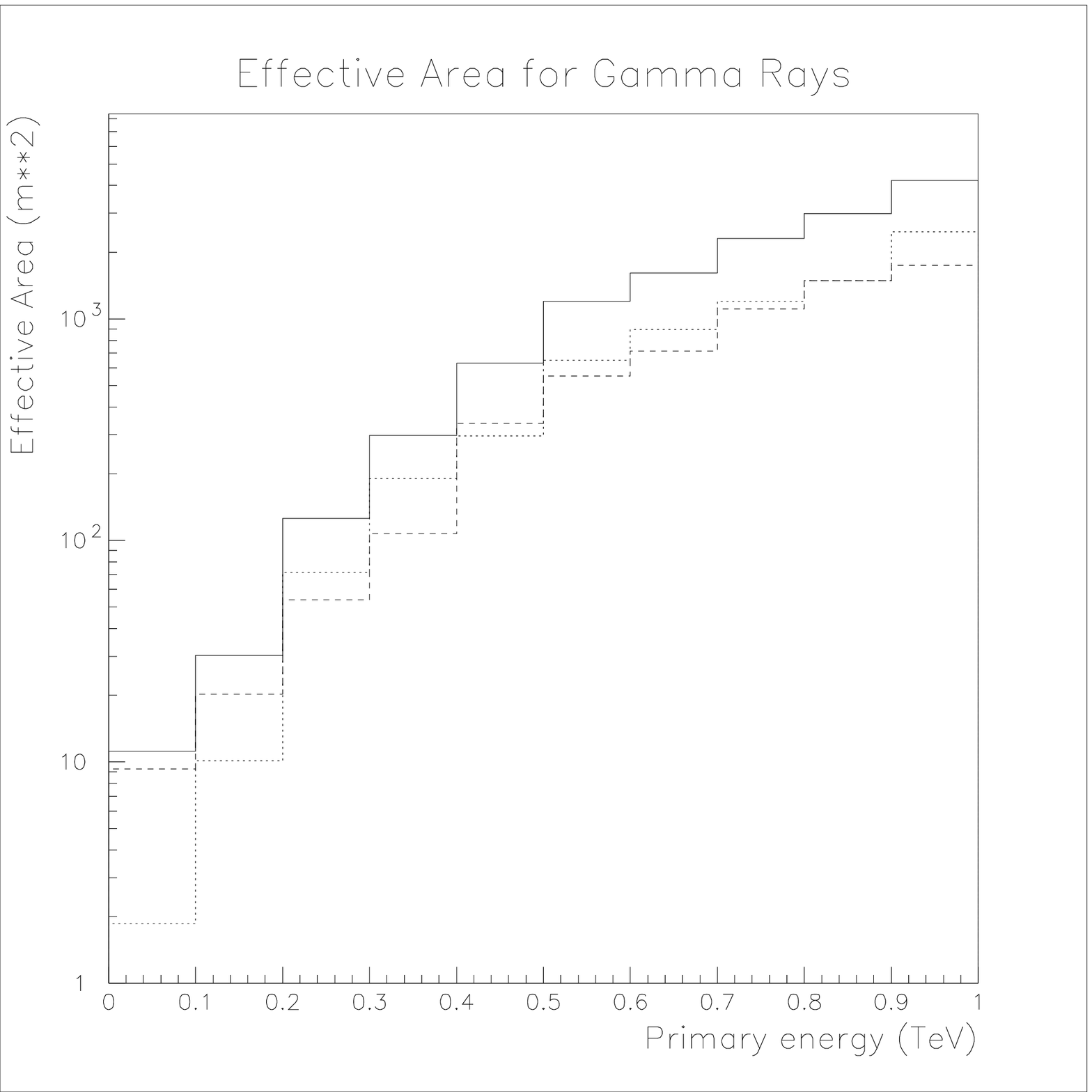,width=2.6in}
\epsfig{file=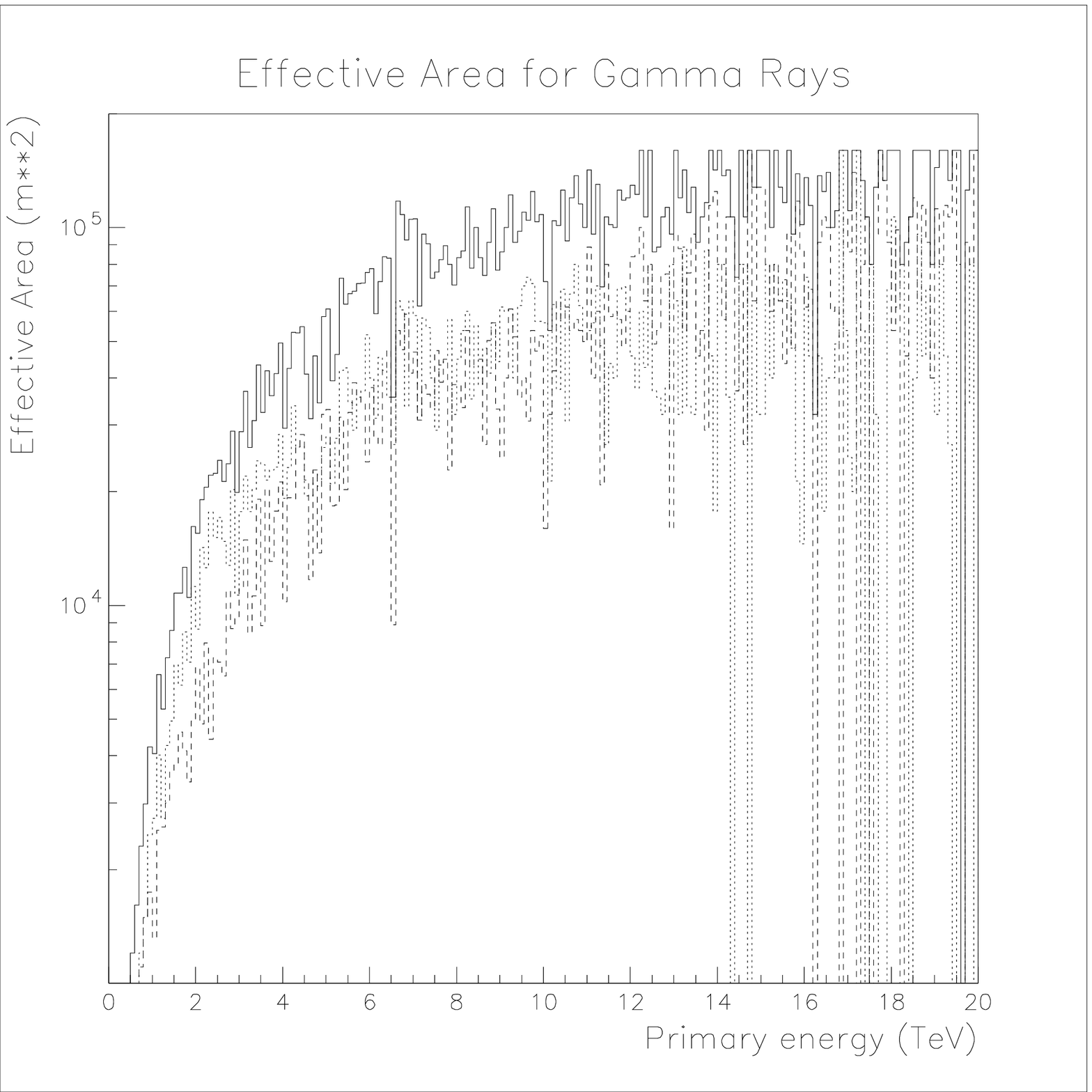,width=2.6in}}}},%
{Plot of effective area versus energy. Dashed is pond-only trigger, dotted is tank-only
trigger, and solid is pond-tank trigger.}]
\end{figwindow}

Those events obtained by using the tanks in a trigger condition have an average pond
pmt multiplicity of 20 and an angular resolution of about $ 2.5^o $. This resolution
is significantly worse than the resolution of pond-triggered events ($ < 1.0^o $) but
is adequate for doing coincident searches with most BATSE-detected Gamma Ray Bursts and
for photon counting analyses where the event angles are not used.

\section{Results from a Prototype Array}

A prototype tank array has been installed near the Milagro pond to study the response
of the water tanks to typical EAS that trigger Milagro.  The array consists of 11
tanks built with commercially available polyethylene storage tanks. The installed pmts
are of the same type as those in the Milagro pond (Hamamatsu R5912).  The tanks are
at various distances from the pond which will enable us to study their response as
a function of EAS core distance and particle density.  The tank hit
multiplicity with at pond trigger requirement of approximately 120 hit pond
pmts is 2.5. Results from these prototypes will also be presented.

\section{Summary}

From the above simulation results one can see the predicted large improvement to
both the angular and core position resolutions of the Milagro detector using information
acquired by a spare array of instrumented Cherenkov water tanks.  This improvement
is mainly for EAS whose cores do not fall directly on the Milagro pond.  Since the
sensitivity of an VHE detector is proportional to its angular precision, this
improvement will have a large positive effect on Milagro's sensitivity.  The greatly
improved core position resolution will increase Milagro's sensitivity to various
source spectral characteristics.

This research was supported in part by the National Science Foundation,
the U. S. Department of Energy Office of High Energy Physics, the U. S. Department
of Energy Office of Nuclear Physics, Los Alamos National Laboratory, the University
of California, the Institute of Geophysics and Planetary Physics, The Research
Corporation, and CalSpace.

%
\vspace{1ex}
\begin{center}
{\Large\bf References}
\end{center}

\noindent
McCullough, J.F. 1999, HE 6.1.02, these ICRC procedings.\\
Westerhoff et al. 1998, 19th Texas Symposium on Relativistic Astrophysics procedings.\\ 
Yodh, G.B. 1996, Space Science Reviews, {\bf 75}: 199-212.\\

\end{document}